\documentclass[useAMS,usenatbib, twocolumn]{mn2e}
\usepackage{graphicx}
\usepackage{natbib}
\include{epsf}

\title[Temperature fluctuations of interstellar dust grains]
{Temperature fluctuations of interstellar dust grains}

\author[K. Horn, H.B. Perets and O. Biham] 
{Kobi Horn$^{1}$, Hagai B. Perets$^{2}$ and Ofer Biham$^{1}$  \\  
$^{1}$ Racah Institute of Physics, The Hebrew University, Jerusalem 
91904, Israel \\
$^{2}$ Faculty of physics, Weizmann Institute of Science, 
Rehovot 76100, Israel}
 
\begin{document}
\maketitle

\begin{abstract}

The temperatures of interstellar dust grains are analyzed 
using stochastic simulations, 
taking into account the grain composition and 
size and the discreteness of the 
photon flux.  
Grains of radii smaller than about 0.02 $\mu$m are found to
exhibit large temperature fluctuations
with narrow spikes following the absorption of UV photons. 
The temperatures of such grains 
may rise by a few $\times$ 10 K for very short times, 
but they do not rise above 80 K even for irradiation
field intensities of photon dominated regions.
The distribution of grain temperatures is 
calculated for a broad range of grain sizes 
and for different intensities of the interstellar radiation field, 
relevant to diffuse 
clouds and to photon dominated regions.
The dependence of the average grain temperature on its
size is shown for different irradiation intensities.
It is found that the average temperatures of grains with radii
smaller than about 0.02 $\mu$m are reduced due to the fluctuations.
The average temperatures of grains of radii larger than
about 0.35 $\mu$m are also slightly reduced due to their more efficient
emission of infrared radiation,
particularly when exposed to high irradiation intensities.
The average temperatures 
$\langle T \rangle$ 
of silicate and carbonaceous grains 
are found to depend on the radiation field 
intensity $\chi_{\rm MMP}$ according to 
$\langle T \rangle \sim {\chi_{\rm MMP}}^{\gamma}$, 
where the exponent
$\gamma$ depends on the grain size and composition.
This fitting formula 
is expected to be useful in 
simulations of interstellar processes,
and can replace commonly used approximations 
which do not account for the grain temperature fluctuations
and for the detailed properties of interstellar dust particles.
The implications of the results on molecular hydrogen formation
are also discussed.
It is concluded that grain-temperature fluctuations tend to reduce
the formation rate of molecular hydrogen,
and cannot account for the observations of H$_2$
in photon-dominated regions, even in the presence of 
chemisorption sites.

\end{abstract}


\section{Introduction}
        \label{intro.}

Interstellar dust grains,
which consist of silicates and carbonaceous materials,
account for about 1\% of the mass
of interstellar clouds.
The grains play an important role in  
processes in the interstellar medium (ISM),
such as the absorption of UV radiation, emission
of infrared radiation 
and heating of the surrounding gas through
the photoelectric effect, which leads to
the emission of energetic electrons.
Catalytic processes on grain surfaces
give rise to the formation of molecular hydrogen
\citep{Gould1963,Hollenbach1971a,Hollenbach1971b}
and other molecules
\citep{Hasegawa1992}.
Dust grains also take part in the dynamics of interstellar
clouds through their role in coupling the magnetic field to the gas
in regions of low fractional ionization and through their transferring
of radiation pressure into the gas 
\citep{Draine2003}. 
Many of these processes depend directly or indirectly 
on various properties of the grains, such as
their temperature and charge. 
These, in turn, depend on both the macro-conditions,
namely the radiation flux and the gas environment,
and the microphysics of the grains namely 
their composition and structure. 
Thus, 
the evolution of interstellar clouds 
involves a complex network of physical
and chemical processes and feedback
mechanisms.
Large scale simulations are required in order to combine
all the known processes and to draw the macro-physical
picture 
\citep{vanHoof2004,LePetit2006}.
An important ingredient 
of interstellar processes is 
the temperature of dust grains and its 
dependence on the interstellar conditions. 
Processes such as infrared emission from the grains 
and gas-grain heating directly affect
the grain temperatures.
Other processes such as the formation of 
H$_2$ and other molecules on grain surfaces 
are highly sensitive to the grain temperatures.

In this paper we calculate the temperatures
of interstellar dust grains and their dependence on
the radiation field 
and on the compositions and sizes of the grains. 
To this end we
perform simulations of stochastic heating
and radiative cooling,
taking into account the discrete nature of the UV photons.
We find that grains of radii smaller than 0.2 $\mu$m
exhibit large temperature fluctuations.
Such small grains exhibit skewed temperature distributions,
where most of the time they are colder than large grains
exposed to similar radiation intensity, but occasionally
they experience a sharp temperature spike. 

The paper is organized as follows.
In section 
\ref{sec:ISRF}, 
we briefly review the 
physical properties of the radiation field 
in different interstellar environments.
Previous studies of interstellar 
grain temperatures are reviewed
in section 
\ref{sec:IDG_temp}. 
The methods used in our stochastic simulations 
are described in section 
\ref{sec:model}. 
The results are presented in section 
\ref{sec:results},
followed by a discussion in section
\ref{sec:discussion}
and a summary in section
\ref{sec:dis_sum}. 

\section{The Interstellar Radiation Field}
\label{sec:ISRF}

\begin{figure}
\includegraphics[width=9cm]{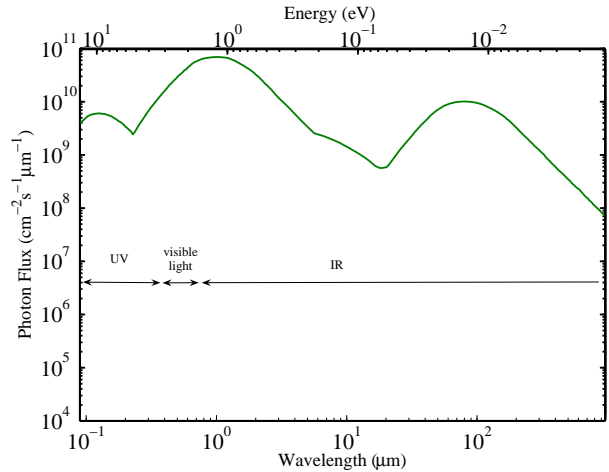}
\caption{
The interstellar radiation field 
$I(\lambda)$,
expressed in terms of the
number of photons per unit wavelength (cm),
per cm$^2$, per second, 
for $\chi_{\rm MMP}=1$.
}
\label{fig:MMP_f_ph}
\end{figure}

The interstellar radiation field
is defined as the average radiation field between stars.
There is a widely accepted approximation
for the typical spectrum 
of the radiation field,
proposed by 
\cite{Mathis1983}.
Within this approximation, the energy density
$u_\lambda$
(eV cm$^{-3}$ $\mu$m$^{-1}$)
of the ambient radiation 
vs. the wavelength $\lambda$
is given by

\begin{equation}
u_{\lambda} = \chi_{\rm  MMP}
\left[u_{\lambda}^{UV_{\odot}} + 
\sum_{i=2}^4W_i{{4\pi}\over{c}}B_{\lambda}(T_i)\right] +
{{4\pi}\over{c}}B_{\lambda}(T_0),
\label{eq:MMP_e}
\end{equation}

\noindent
where
$\chi_{\rm MMP}$
is the enhancement of the starlight component relative to
the solar neighborhood
\citep{Habing1968,Mathis1983}
and $c$ (cm s$^{-1}$)
is the speed of light. 
The term
$u_{\lambda}^{UV_{\odot}}$ 
accounts for the UV component of the
solar neighborhood.
Its explicit form is given in
\cite{Mezger1982}.
The parameters are
$W_2=10^{-14}$,
$W_3=10^{-13}$,
$W_4=4\times10^{-13}$,
$T_2=7500$K,
$T_3=4000$K
and
$T_4=3000$K,
while 
$T_0=2.7$K is the temperature of the 
cosmic background radiation.
The function 
$B_{\lambda}(T)$ 
is the Planck distribution for 
black-body radiation
at temperature $T$ (K)
given by

\begin{equation}
B_{\lambda}(T) = 
{{2hc^2}\over{\lambda^5}}
{\left({e^{{hc}\over{{\lambda}k_B T}}-1}\right)^{-1}},
\label{eq:Planck_function}
\end{equation}

\noindent
where $h$ 
(eV s)
is the Planck constant.
The photon flux 
$I(\lambda) = \lambda u_{\lambda}/h$
(photons cm$^{-2}$ s$^{-1}$ $\mu$m$^{-1}$),
obtained from 
Eq. (\ref{eq:Planck_function}) for $\chi_{\rm MMP}=1$,
is shown in 
Fig. \ref{fig:MMP_f_ph},
for the relevant range of wavelengths 
$0.0912 < \lambda < 975$ $\mu$m
(or photon energies in the range
$1.27 \times 10^{-3} <E< 13.6$ eV).
Photons that are more energetic than $13.6$ eV
are suppressed inside the cloud because they
ionize hydrogen atoms on the edge of
the cloud and cannot penetrate further.

\section{Dust-grain temperatures}
\label{sec:IDG_temp}

Information on the properties of interstellar
dust grains is obtained from spectroscopic 
observations of the extinction, scattering and emission
of electromagnetic radiation by these grains.
Such observations show that the grains consist of silicate and
carbonaceous materials.
Silicate grains consist of 
(Mg$_x$Fe$_{1-x}$)$_2$SiO$_4$,
where
$0<x<1$.
Carbonaceous grains include
crystalline graphitic carbon, amorphous carbon and 
polycyclic aromatic hydrocarbons (PAHs).
In cold molecular clouds the grains are coated by
ice mantles.
However, here
we focus on bare silicate
and carbon grains,
relevant to diffuse clouds and photon-dominated regions.

The size distribution of interstellar dust 
grains follows an approximate power law
of the form
$dn(r)/dr \propto r^{-\lambda}$, 
where $r$ is the grain radius,
$n(r)$ (cm$^{-3}$) is the density of grains smaller than $r$ 
and
$\lambda = 2.5$.
Such distribution, between suitable cutoffs 
$r_{\rm min} < r < r_{\rm max}$, 
was shown to reproduce the observed extinction curve 
\citep{Mathis1977}. 
Later work by Draine \& Lee (1984)
provided further evidence for the model. 
However a more recent study by 
Weingartner \& Draine (2001)
has shown that a range of different
size distributions exist compatible with observations of
different regions in the Milky Way, the LMC and the SMC galaxies. 

The temperatures of interstellar dust grains
depend on the heating and cooling processes
due to the interaction of these grains
with the radiation fields and the surrounding gas.
These processes include the absorption and
emission of electromagnetic radiation, gas-grain
collisions, heating by cosmic rays and surface
reactions.
In most interstellar environments, the grain heating and
cooling processes are dominated
by absorption and emission of radiation.
Gas-grain collisions strongly affect grain temperatures only in dense 
regions where the rate of such collisions is high
\citep{Spitzer1978}. 
Chemical reactions affect the grain temperatures
in very cold dense regions in the cores of molecular clouds.

Dust grain temperatures
have been studied for many years.
DeHulst (1949)
estimated the temperature of dielectric particles to be
$\sim 15$K.
Detailed 
calculations of the temperatures of grains of different 
compositions in different
radiation fields were done by
\cite{Greenberg1971}.
Later, a graphite-silicate model 
was proposed by
\cite{Mathis1977}. 

The temperature fluctuations of small grains
were analyzed by
\cite{Draine1984} 
and by 
\cite{Dwek1986},
using statistical methods.
These focused mainly on the infrared emission
from such grains.
Li \& Draine (2001) 
evaluated the temperatures of
dust grains as a function of their sizes and
the ambient radiation field, using the balance between
the absorbed and emitted radiation:

\begin{equation}
\int_{0}^{\infty} 
{ Q_{\rm abs}(r,\lambda)cu_{\lambda}d{\lambda}} 
= \int_ {0}^{\infty} 
{4\pi Q_{\rm abs}(r,\lambda) B_{\lambda}(T)d{\lambda}}.
\label{eq:Thermal_equilibrium_temperature}
\end{equation}

\noindent
Here, 
$\pi r^2 Q_{\rm abs}(r,\lambda)$ 
is the absorption cross-section
for a photon of wavelength 
$\lambda$ (cm) 
by a grain of radius $r$
where $Q_{\rm abs}(r,\lambda)$
is the absorption coefficient.
The energy density vs. wavelength,
$u_{\lambda}$ (eV cm$^{-3}$ $\mu$m$^{-1}$), 
is given by Eq. 
(\ref{eq:MMP_e}).
Thus,
$c u_\lambda$ (eV cm$^{-2}$ s$^{-1}$ $\mu$m$^{-1}$) 
is the energy flux density at wavelength 
$\lambda$.
Eq. (\ref{eq:Thermal_equilibrium_temperature})
describes the balance between
absorption (left-hand side) and emission 
(right-hand side)
for large grains that maintain a nearly
constant temperature ${T}$ (K).
While most of the energy absorbed by grains is
in the UV range, the emitted radiation is in the
infrared range.
From Eq.
(\ref{eq:Thermal_equilibrium_temperature})
one can extract the grain temperature $T$
vs. the grain size $r$ and the radiation 
field intensity, given by
$\chi_{\rm MMP}$.
Note that the temperatures of such large grains
do not depend on the heat capacity of the grain.

\section{The Stochastic Simulations}
\label{sec:model}

Here we present
stochastic simulations of grain heating and cooling,
taking into account the grain 
composition and size and the discrete
nature of the UV photons.
In the simulations we follow the temporal 
evolution of the temperature of a single grain.
The spectrum of absorbed photons is divided into two domains:
(a) the absorption of an energetic UV 
photon is taken as a discrete, stochastic process. 
(b) The contribution of the less energetic photons is
considered as a continuous flux of energy.
The emission of infrared radiation by the grains, 
is also described as a continuous process.

The time-dependent temperature of a single grain 
is evaluated as follows.
Starting from a given initial temperature, the grain
temperature is updated taking into 
account both the discrete and the continuous 
processes.
The continuous variation of $T(t)$ 
is calculated by integrating the 
absorbed and emitted radiation, 
taking into account the
heat capacity of the grain.
To evaluate the stochastic contribution we
specify the times in which UV photons
are absorbed by the grain.
These times are
drawn from a suitable Poisson distribution,
which accounts for the 
flux
of energetic photons and the
wavelength-dependent cross
section of the grain.
For each absorption event,
the wavelength of the photon is 
drawn from a suitable distrituion obtained from
$u_{\lambda}$
and
$Q_{\rm abs}$.
The grain temperature is then updated
according to the energy of the photon and the
temperature-dependent heat capacity of the grain.

\subsection{Physical Properties of Dust Grains}
\label{sec:phys_prop}

\begin{figure}
\includegraphics[width=9cm]{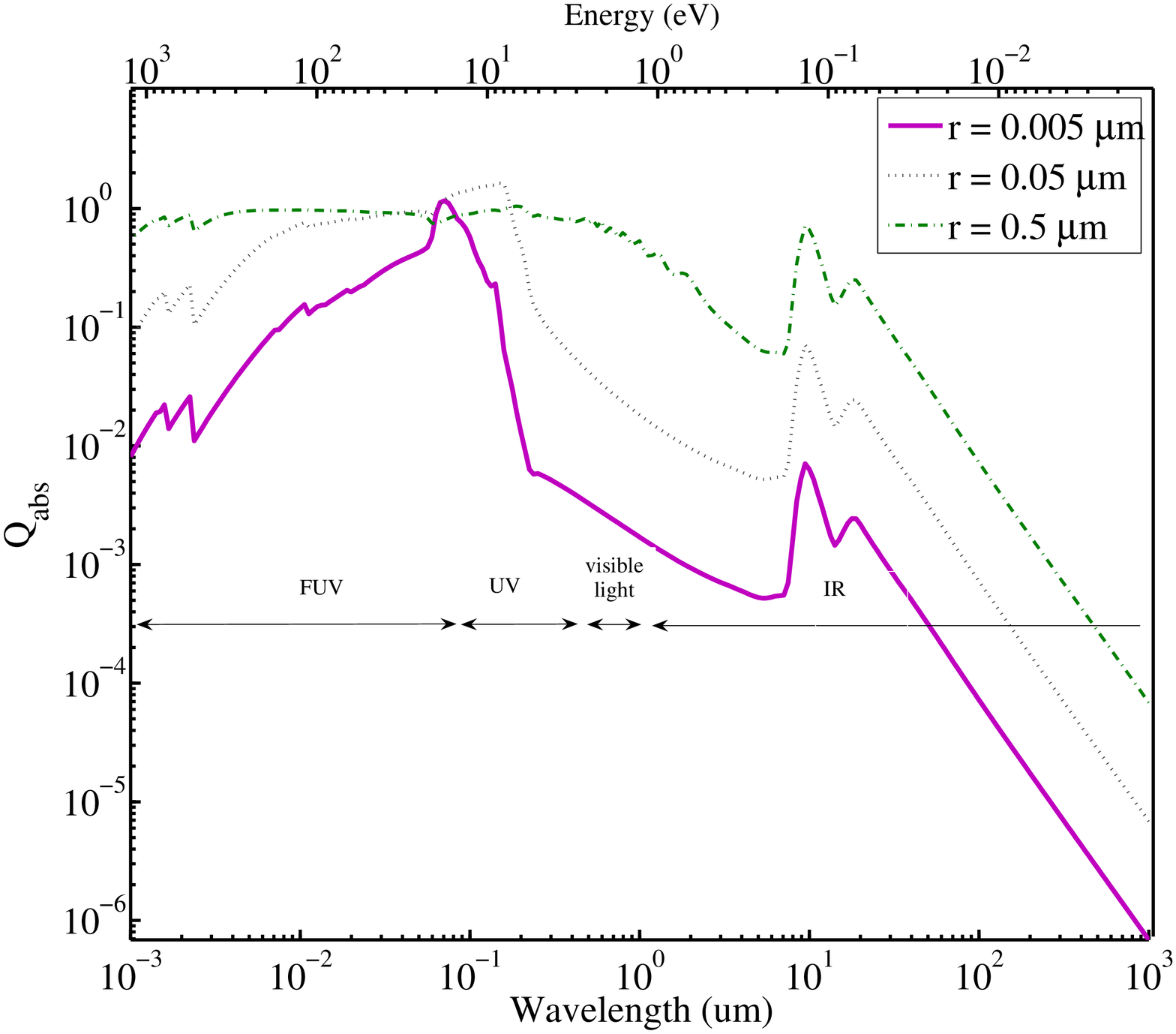}
\caption{
The cross-section $Q_{\rm abs}(\lambda)$
for photon absorption onto a graphite grain, 
as a function of the wavelength.
}
\label{fig:Qabs}
\end{figure}

For simplicity, we assume that the grains are
spherical and denote the grain radius by $r$.
Two grain compositions are studied:
carbon (or graphite) grains with mass density of 
$2.16$ (gram cm$^{-3}$) 
and silicate grains with mass density of 
$3.5$ (gram cm$^{-3}$).
In the analysis we use the physical properties of such
spherical, single-component grains given by
Laor \& Draine (1993). 
The photon absorption coefficient 
$Q_{\rm abs}$
as function of the wavelength 
${\lambda}$ 
for graphite grains
is shown in
Fig. \ref{fig:Qabs}.
This absorption coefficient depends on the grain size.
It is also weakly dependent on the temperature,
however this dependence is neglected in our analysis.
The values of
$Q_{\rm abs}$
used here are for grain temperature of 25 K
(Laor \& Draine 1993).

\begin{figure}
\includegraphics[width=9cm]{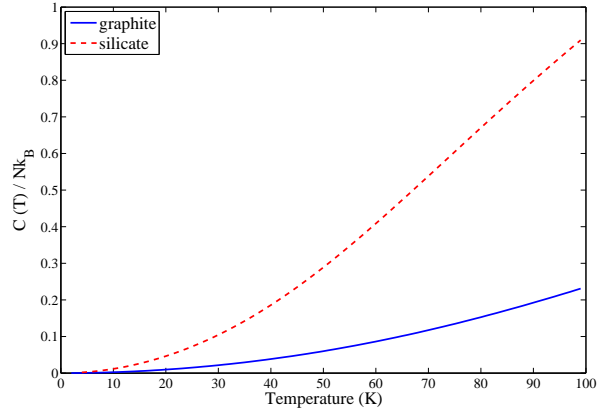}
\caption{
The heat capacity per carbon atom in the graphite grain (solid line) and
per atom in the silicate grain (dashed line) vs. grain temperature.
Here $C(T)$ stands for $C_{graphite}(T)$ or $C_{silicate}(T)$, 
while $N$ is $N_c$ or $N_a$ accordingly
(Draine et al. 2001). 
the Boltzmann constant 
k$_B$ is 
given in units of eV/K.
}
\label{fig:Cv_T}
\end{figure}

The heat capacities of silicate and graphite grains 
are shown in
Fig. \ref{fig:Cv_T}. 
The heat capacity of a graphite grain at temperature $T$ 
can be expressed by
\citep{Draine2001}

\begin{equation}
C_{\rm graphite}(T) = 
(N-2)k_B \left[f'_2 \left ( {T\over 863}\right)
 + 2f'_2  \left ( {T\over 2504}\right) \right],
\label{eq:heat capacity graphite}
\end{equation}

\noindent
where $N$ is the number of carbon atoms in the grain and 
$f'_n(x) \equiv df_n(x)/dx$,
where

\begin{equation}
f_n(x)\equiv n{\int_ {0}^{1} {y^{n}dy\over {{\rm exp}(y/x)-1}}}.   
\label{eq:heat capacity help function}
\end{equation}

\noindent
Similarly, the heat capacity of a silicate grain is given by
\citep{Draine2001}

\begin{equation}
C_{\rm silicate}(T)
= (N - 2 )k_B \left[2f'_2 \left ( {T\over 500}\right)
 + f'_3  \left ( {T\over 1500}\right) \right],
\label{eq:heat capacity silicate}
\end{equation}

\noindent
where $N$ is the number of atoms in the grain.

\subsection{The Photon Flux}

In order to simulate the time dependent
temperature of a grain exposed to a
given radiation field,
we first divide the relevant spectrum of the radiation field
into two domains: the hard, energetic photons,
taken discretely, and the soft photons,
taken as a continuous flux.
For the discrete part 
we evaluate the average time $\tau$ between absorptions
of hard photons.
The actual absorption times are then obtained, using
a poisson distribution, 
with an average $\tau$,
for the times between absorption events.
The wavelength of each absorbed photon is drawn from
a suitable distribution,  
which accounts for the interstellar radiation field
and the absorption coefficient 
$Q_{\rm abs}$.

The threshold 
${\lambda_c(r)}$ 
between the domains 
of hard and soft photons
depends on the grain radius $r$.
The criterion for hard photons is that these
photons are sufficiently energetic to cause a significant
change in the grain temperature upon absorption.
In practice, 
${\lambda_c(r)}$ 
is determined such that photons with
$\lambda < \lambda_c$ 
raise the temperature of a grain of radius $r$
from 
$T_0$ 
to 
$T > T_0 + \Delta T$. 
More specifically
we use 
$T_0 = 10$ K 
and 
$\Delta T = 0.005$ K.
These values provide a good description
of the stochastic features while maintaining a reasonable
running time of the simulation even for large grains.

\subsection{Stochastic Heating of Grains}

The stochastic heating of a grain of radius $r$ exposed to radiation
field $u_{\lambda}$ is simulated as follows.
First, we evaluate the average time, $\tau$,
between successive absorptions of energetic photons.
This time is given by 
\citep{Draine1985}

\begin{equation}
\tau = \left({\pi r^2 
\int_{\lambda_{min}}^{\lambda_c(r)} 
{I(\lambda) Q_{\rm abs}(r,{\lambda}) d\lambda}}\right)^{-1},
\label{eq:Tau}
\end{equation}

\noindent
where 
$\lambda_{\rm min} = 0.0912$ ($\mu$m) 
is the lower cutoff of the spectrum, 
which corresponds to ionization threshold 
of atomic hydrogen.
The times, 
$\Delta t$, 
between photon absorption events are 
drawn from the distribution 
$P(\Delta t) = \exp({-\Delta t/\tau})/\tau$.
The next step is to obtain the wavelength (and thus the energy) of the
absorbed photon.
This wavelength is drawn from the distribution

\begin{equation}
P(\lambda) = \frac{I(\lambda) Q_{\rm abs}(r,\lambda)}
{\int_{\lambda_{min}}^{\lambda_c(r)} { I(\lambda) 
Q_{\rm abs}(r,{\lambda}) d\lambda}}.
\label{eq:PofLambda}
\end{equation}

The temperature T 
of a grain immediately after the absorption of a UV photon
of energy $E_{\rm photon} = h c/\lambda$
is calculated using the relation

\begin{equation}
E_{\rm photon} = 
\int_{T_0}^{T} C(T')dT',
\label{eq:final_temp}
\end{equation}

\noindent
where $C(T')$ is the heat capacity of the grain and
$T_0$ is the instantaneous grain temperature 
before the absorption of the photon.
To avoid an accumulation of numerical errors,
the temperature $T$ is calculated analytically.
To this end, the heat capacity $C(T)$ is fitted 
by a polynomial function, which is then integrated,
giving rise to an expression of $T$ in terms of 
$T_0$ and $E_{\rm photon}$.

\subsection{Continuous Heating and Radiative Cooling of Grains}

The time dependence of the grain temperature 
between absorptions of UV photons
is determined by the balance between the
absorption of soft photons and the
emitted radiation.
The time derivative of the grain temperature is thus
given by

\begin{equation}
{dT \over dt} = 
{3\over{rC(T)}}
\left[H - 4\pi\int_{0}^{\infty}
{Q_{abs}({\lambda})B_\lambda(T)d\lambda}\right],
\label{eq:Temperature_equation}
\end{equation}

\noindent
where

\begin{equation}
H(r) \equiv {c\over4} 
\int_{\lambda_c(r)}^{\infty}
{Q_{abs}({\lambda})u_\lambda d\lambda}
\label{eq:Soft_photons}
\end{equation}

\noindent
is the radiation absorbed by a grain of radius $r$
within the frequency range of soft photons.  
To obtain the time-dependent grain temperature $T(t)$,
Eq. (\ref{eq:Temperature_equation}) 
is integrated using a
standard Runge Kutta stepper
during the period between succesive absorption
events of hard photons.

\section{Results}
\label{sec:results}

\subsection{Grain Temperature vs. Time}
\begin{figure}
\includegraphics[width=9cm]{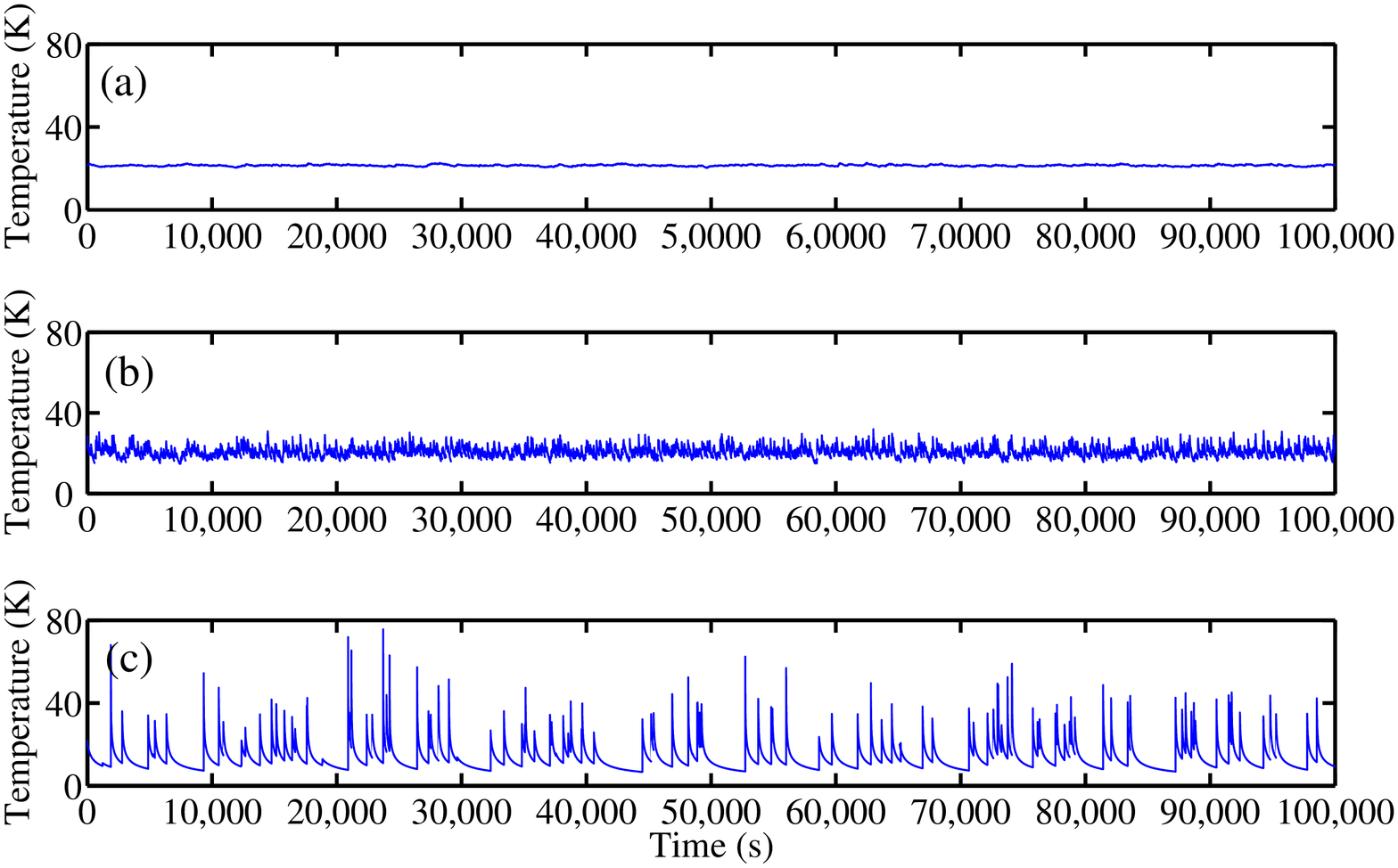}
\caption{
The temperature variations of a {\bf carbon} grain during one day 
in which it is exposed radiation intensity of 
$\chi_{\rm MMP}=1$,
where the grain radii are:
(a) 0.05; 
(b) 0.016;
and 
(c) 0.005 
$\mu$m.
}
\label{fig:Tvt_gra_1}
\end{figure}

\begin{figure}
\includegraphics[width=9cm]{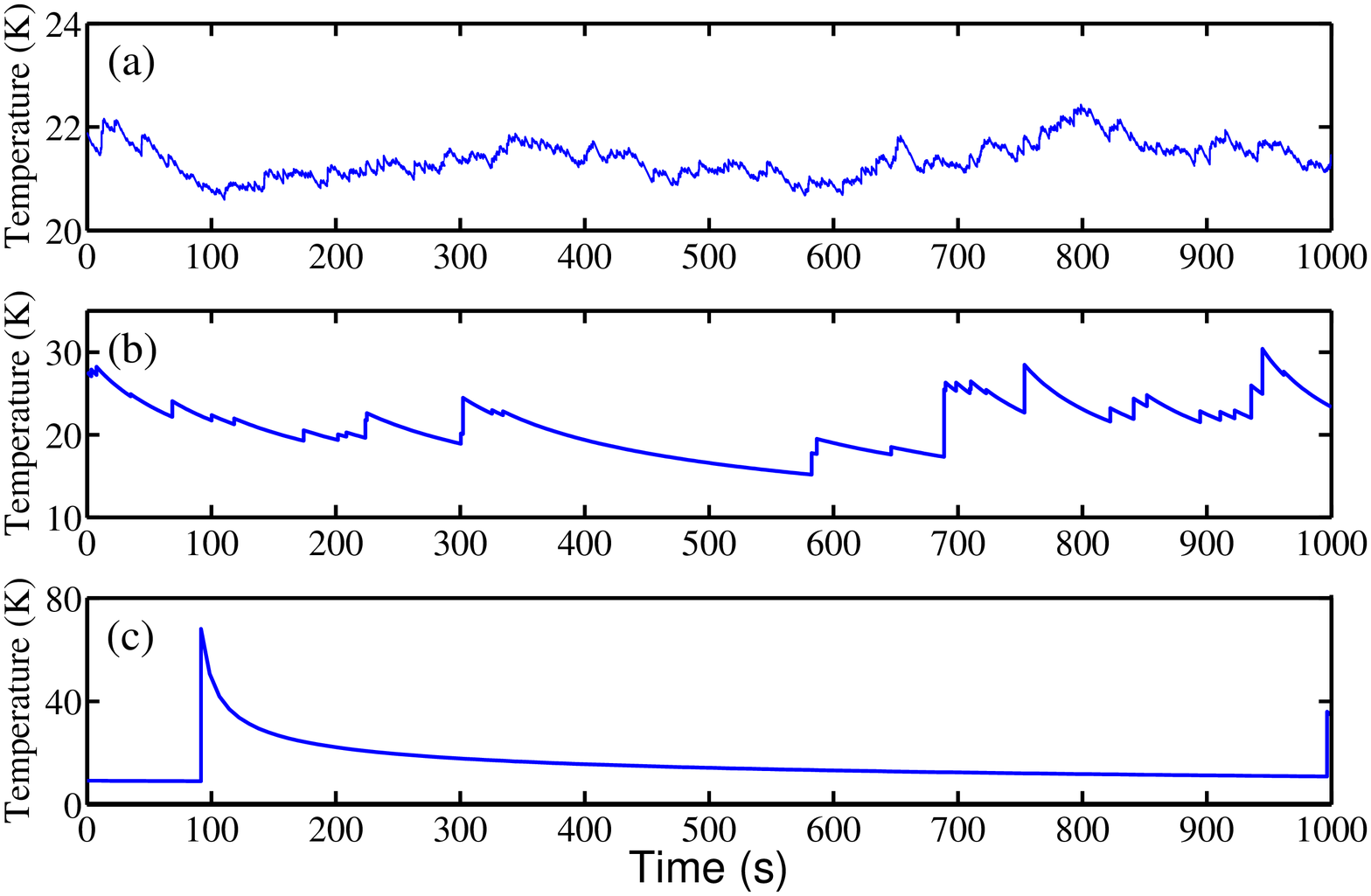}
\caption{
The temperature variation of the three {\bf carbon} grains
shown in Fig.
\ref{fig:Tvt_gra_1}
during 1000 seconds in which they are exposed to radiation 
intensity of  
$\chi_{\rm MMP}=1$.
}
\label{fig:Tvt_sil_1}
\end{figure}

In
Fig. \ref{fig:Tvt_gra_1} 
we present the grain temperature 
vs. time for carbon 
grains
with radii
of (a) 0.05, (b) 0.016 and (c) 0.005 $\mu$m
for radiation intensity of $\chi_{\rm MMP}=1$.
Clearly, for the largest grain size 
the temperature fluctuations
are negligible while for the smallest grain they
are intense.
The temperature of the small grains is low most of the time 
(10-15K).
However, following the absorption of a UV photon,
the grain temperature
may rise up to nearly 80K
followed by fast cooling,
giving rise to narrow temperature
spikes. 
This is due to the fact that the emitted radiation intensity 
sharply increases as a function of
the grain temperature.
As a result, the fluctuations tend to reduce the average temperature
of the small grains.
In Fig. 
\ref{fig:Tvt_sil_1}
we present the time dependent grain temperatures with a higher
temporal resolution. 
The dependence of the rate of UV photon absorptions on the
grain size is clearly observed.

\subsection{Grain-Temperature Distributions}

Due to the wide variations in the temperatures of small
grains, one should calculate not only the average temperature
$\langle T \rangle$
but the entire probability density function
$f(T)$.
This density function is
obtained directly from $T(t)$
by calculating what fraction of time the grain has
spent in each range of temperatures. 
\begin{figure}
\includegraphics[width=9cm]{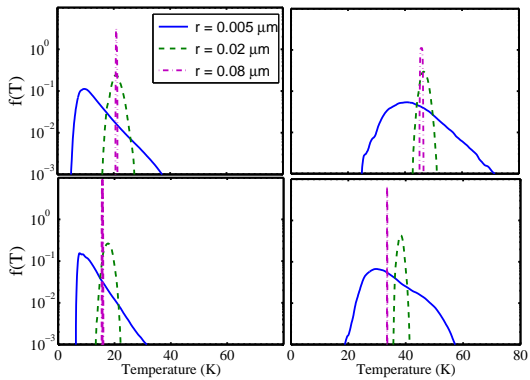}
\caption{
The distributions of grain tempeatures for grains
of radii of 0.005 (solid line), 0.02 (dashed line) 
and 0.08 (dashed-dotted line) $\mu$m:
(a) carbon grain exposed to radiation of $\chi_{\rm MMP}=1$;  
(b) carbon grain exposed to $\chi_{\rm MMP}=100$;
(c) silicate grain exposed to $\chi_{\rm MMP} = 1$;
and
(d) silicate grain exposed to $\chi_{\rm MMP} = 100$.
}
\label{fig:T_dis_1}
\end{figure}
In 
Fig. \ref{fig:T_dis_1}
we present the distribution of grain temperatures  
for carbon
grains
[Figs. \ref{fig:T_dis_1}(a) and \ref{fig:T_dis_1}(b)]
and for silicate grains
[Figs. \ref{fig:T_dis_1}(c) and \ref{fig:T_dis_1}(d)].
The results are shown for three grain radii,
5 (solid line), 20 (dashed line) and
80 nm (dashed-dotted line)
and two radiation field intensities: $\chi_{\rm MMP}=1$
[Figs. \ref{fig:T_dis_1}(a) and \ref{fig:T_dis_1}(c)]
and $\chi_{\rm MMP}=100$
[Figs. \ref{fig:T_dis_1}(b) and \ref{fig:T_dis_1}(d)].
The large grains exhibit very narrow temperature 
distributions.
This is due to the fact that the large grains absorb a high
flux of UV photons and that the effect of each photon
on their temperature is small.
As the radiation field increases the  
peak shifts to higher temperatures.
The small grains exhibit broad temperature distributions.
These distributions are skewed, where
most of the weight is concentrated at the
low temperature side with only a narrow 
tail at the high temperature side.
The typical temperature of the small grains
is lower by about $10$ K compared
to the large grains.

\subsection{Average Grain Temperatures}
\begin{figure}
\includegraphics[width=9cm]{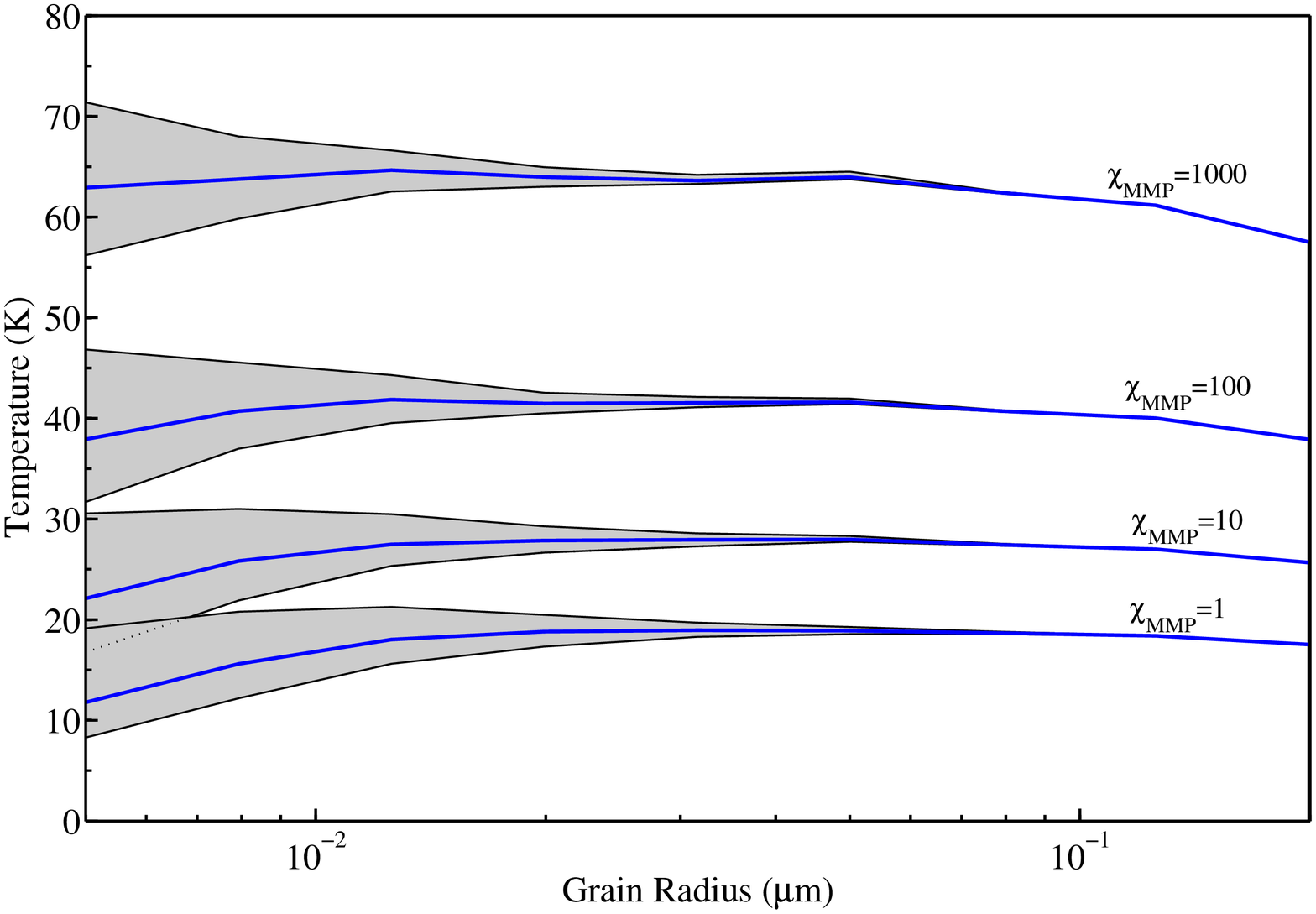}
\caption{
The average temperature and fluctuation range vs. grain 
radius for carbon grains exposed to different
irradiation field intensities. 
}
\label{fig:T_vs_R_gra}
\end{figure}
\begin{figure}
\includegraphics[width=9cm]{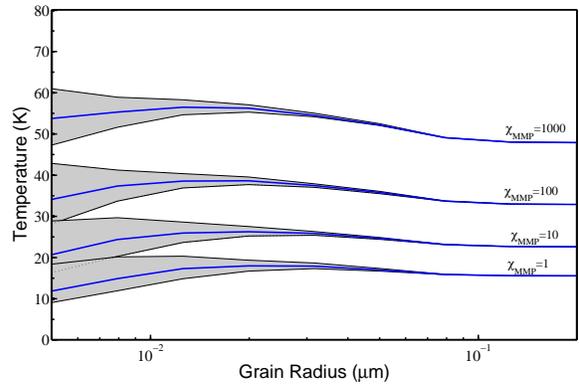}
\caption{
The average temperature and fluctuation range vs. grain 
radius for silicate grains exposed to different
irradiation field intensities. 
}
\label{fig:T_vs_R_sil}
\end{figure}

In Figs. 
\ref{fig:T_vs_R_gra} 
and 
\ref{fig:T_vs_R_sil}
we present the average grain temperatures 
$\langle T \rangle$
(solid lines)
vs. grain radius for 
carbon and silicate grains,
respectively,
for radiation intensities
of $\chi_{\rm MMP}=1$, 10, 100 and 1000.
The gray areas represent the range of variation of the
grain temperature. 
The upper (lower) bounds, 
$T_{+}(r)$ ($T_{-}(r)$) of these areas represent the
upper (lower) semi-standard deviations with respect to 
$\langle T \rangle$,
computed from the 
probability density function
$f(T)$, taking into account
only the range 
$T > \langle T \rangle$ ($T < \langle T \rangle)$.
More specifically,

\begin{equation}
T_{+}(r) = \langle T \rangle 
+ \left[ 
\frac{\int_{\langle T \rangle}^{\infty} (T - \langle T \rangle)^2 f(T)}
{\int_{\langle T \rangle}^{\infty} f(T)} 
\right]^{1/2}
\end{equation}

\noindent
and

\begin{equation}
T_{-}(r) = \langle T \rangle 
- \left[
\frac{\int_{0}^{\langle T \rangle} (T - \langle T \rangle)^2  f(T)}
{ \int_{0}^{\langle T \rangle} f(T)}
\right]^{1/2}.
\end{equation}

\noindent
It is found that for all radiation fields, grains of
radii larger than 
$r_c = 0.02$ $\mu$m 
do not exhibit 
significant temperature fluctuations. 
For grains of radii $r < r_c$, the temperature fluctuations are
enhanced as $d$ is reduced.
The threshold value of the diameter below which fluctuations
are significant is clearly independent of the radiation field.
This is due to the fact that the fluctuation level is determined
by the temperature rise that is caused by a single UV photon,
which is reduced as the grain mass is increased.
For small grains, the temperature spikes caused by UV photons are very narrow
and exhibit little overlap even for high radiation fields.
For high radiation fields the average temperature, $\langle T \rangle$
(solid line) 
is at about the middle of the gray area. For small grains,  
the average temperature is near the lower edge of the gray area. 
This is due to the fact that the small grains spend most of the time
at low temperatures.

The results for 
the temperature distributions
(Fig. 6)
and for the average grain temperatures 
(Figs. 7 and 8),
for $\chi_{\rm MMP}=1$,
are in agreement with 
Li \& Draine (2001)
and with 
Cuppen, Morata \& Herbst (2006)
in the corresponding range of grain sizes,
up to variations of the order of 1K.
The results presented here extend the
analysis of grain temperatures 
to higher radiation fields.
These results can thus be used in modeling of both
diffuse clouds and photon-dominated regions.

\begin{figure}
\includegraphics[width=9cm]{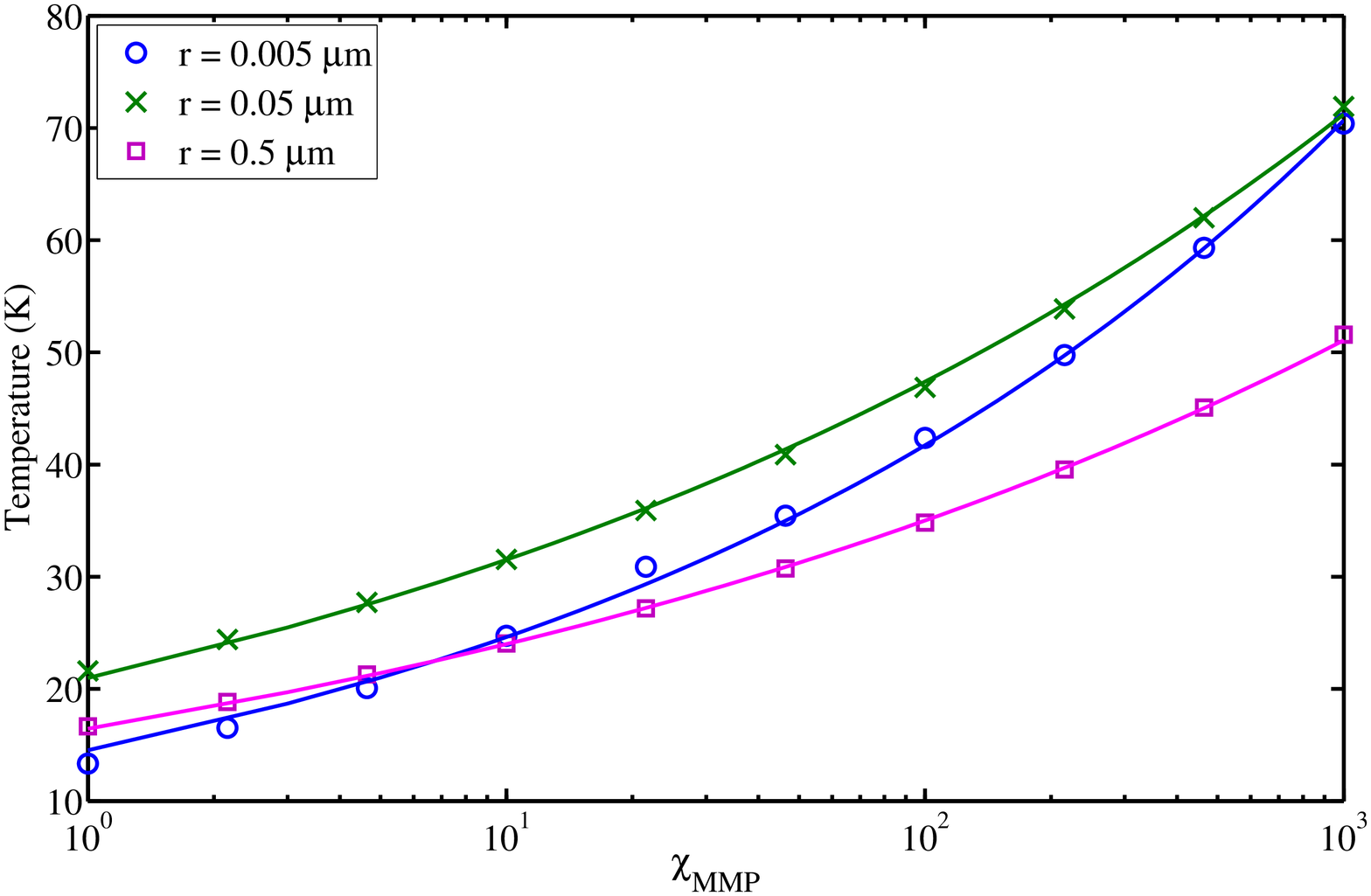}
\caption{
The average temperature vs. interstellar radiation field $\chi_{\rm MMP}$
for carbon grains with radii of
0.005 (solid line), 0.05 (dashed line) and
0.5 (dashed-dotted line) $\mu$m.
}
\label{fig:T_vs_F_gra}
\end{figure}

\begin{figure}
\includegraphics[width=9cm]{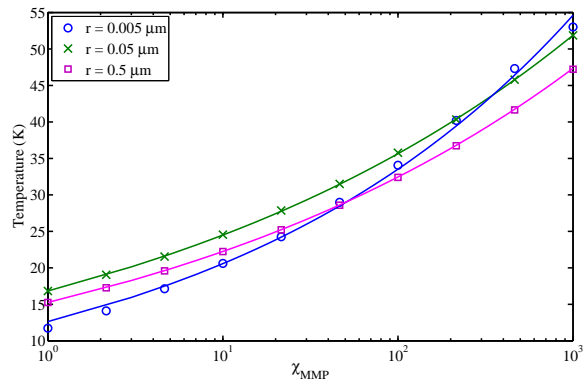}
\caption{
The average temperature vs. interstellar radiation field $\chi_{\rm MMP}$
for silicate grains with radii of
0.005 (solid line), 0.05 (dashed line) and
0.5 (dashed-dotted line) $\mu$m.
}
\label{fig:T_vs_F_sil}
\end{figure}

The highest average temperatures are obtained 
for grains of
intermediate size, 
with radii in the range of 
20-35 nm. 
Smaller and larger 
grains exhibit lower average temperatures.
This can be understood as follows.
Small grains radiate much
more efficiently during the
short times they spend at high temperatures and 
thus tend to have lower average 
temperatures.
Also, very small grains absorb UV radiation less
efficiently than large grains,
in both graphite (Fig. 2) and silicate grains. 
Large grains radiate much more efficiently 
in the relevant range of infrared radiation.
The average temperatures of grains of different sizes as a function of
the radiation field intensity $\chi_{\rm MMP}$ are shown
in Figs. 
\ref{fig:T_vs_F_gra} 
and 
\ref{fig:T_vs_F_sil} . 
These curves are fitted by the function 

\begin{equation}
\langle T \rangle = K {\chi_{\rm MMP}}^{\gamma},
\label{eq:T_f}
\end{equation}

\noindent
where $K$ and $\gamma$ 
depend on the grain size.
The values of $K$ and $\gamma$
for grains of three different sizes,
which consist of
silicates and carbon 
are shown in Tables
1 and 2.

\begin{table}
\caption{
Fitting parameters for the average grain temperature 
vs. radiation field
for carbon grains 
\label{t:E_barriersC} 
}

\begin{centering}\begin{tabular}{ccc}
\hline
Grain Radius ($\mu$m) & $K$ & $\gamma$ \\
\tabularnewline
\hline
0.005 & 14.5 & 0.23 \\
0.05  & 21.0 & 0.18 \\
0.5   & 16.4 & 0.16
\tabularnewline
\hline
\end{tabular}\par\end{centering}
\end{table}

\begin{table}
\caption{
Fitting parameters for the average grain temperature 
vs. radiation field
for silicate grains 
\label{t:E_barriersS}
} 

\begin{centering}\begin{tabular}{ccc}
\hline
Grain Radius ($\mu$m) & $K$ & $\gamma$ \\
\tabularnewline
\hline
0.005 & 12.6 & 0.21 \\
0.05  & 16.8 & 0.16 \\
0.5   & 15.2 & 0.16 
\tabularnewline
\hline
\end{tabular}\par\end{centering}
\end{table}

To interpret the resulting values of $\gamma$,
consider first a macroscopic object which absorbs
and emits radiation like a black body.
For such object
$Q_{\rm abs} \equiv 1$.
Its temperature $T$ does not exhibit fluctuations.
Since the intensity of the radiation emitted from
such object is proportional to $T^4$, 
one obtains that $\gamma = 1/4$.
Unlike a black body, for large interstellar grains,
in the infrared regime, 
$Q_{\rm abs} \sim \lambda^{-2}$.
According to Wien's law,
for an object at temperature $T$,
the highest radiation flux is emitted at
wavelength
$\lambda_{\rm max}(T)=w/T$, 
where
$w=0.29$ (cm K)
is the Wien's constant.
As a result, the emitted radiation is proportional
to 
$Q_{\rm abs}(\lambda_{\rm max}(T)) T^4 \sim T^6$,
while the incoming radiation is proportional to
$\chi_{\rm MMP}$.
Thus, $T \sim {\chi_{\rm MMP}}^{1/6}$,
namely $\gamma = 1/6$.
This analysis agrees very well with the 
results shown in Tables 1 and 2 for silicate
and carbon grains of radii
$r=0.5$ $\mu$m. 
For very small grains, stochastic fluctuations
reduce the average temperature in the limit
of low radiation intensity.
As a result, larger values of $\gamma$ are obtained.

\section{Discussion}
\label{sec:discussion}

The results presented above are expected to be useful 
in simulations of processes taking place in interstellar
clouds. In particular, the average grain temperatures
vs. grain size and radiation intensity can be used as
input data for such simulations.
Also, the grain-temperature distribution, combined with
the grain size distribution can be used to evaluate the
infrared radiation spectrum emitted from dust grains in 
an interstellar cloud.

The commonly used approach to the evaluation of grain
temperatures are based on the balance between the
absorbed and emitted radiation, given by 
Eq. (3).
Using Eq. (3) as an implicit equation, the
grain temperature can be extracted numerically.
In this case one can take into account the
dependence of the absorption and emission
properties on the grain size, through the
function $Q_{\rm abs}$.

In this approach 
the stochastic heating which leads to
temperature fluctuations of small grains is
ignored.
Note that due to the broad distribution
of grain sizes, most of the surface area
of interstellar dust is in small grains.
Thus small grains dominate the radiation
emitted from grains.
In order to obtain an explicit formula
for the grain temperature, the function 
$Q_{\rm abs}$ is often replaced by a crude approximation,
in which it is constant in the UV range, and decays
as $1/\lambda$ in the infrared range
[see Eq. (2) in
\cite{Hollenbach1991}].
This approximation ingnores the specific features
of the silicate and carbon materials as well as
the dependence of
$Q_{\rm abs}$
on the grain size.
For large grains, the assumption of 
$Q_{\rm abs}=1$
in the UV range is valid.
However, in Fig. 2 it appears that in the 
range of wavelengths in which most of the 
radiation from grains takes place
($\lambda > 30$ $\mu$m)
the functional dependence is
$Q_{\rm abs} \sim 1/\lambda^2$
rather than $1/\lambda$ as assumed in
\cite{Hollenbach1991}].
This feature directly affects the dependence of the
average grain temperature on the radiation field
intensity
(see Sec. 5.3 above).
For small grains, the assumption of constant
$Q_{\rm abs}$
is not valid. 
More specifically,
for small grains
$Q_{\rm abs}$
strongly depends on $\lambda$
and on the grain size
(Fig. 2).

Our analysis takes into account both the specific features
of the silicate and carbon materials, including the detailed
dependence of the absorption and emission on the wavelength
and on the grain size. 
In addition, the effect of stochastic heating is taken into
account.
This approach provides a more precise evaluation of interestellar
dust temperatures under a broad range of conditions.
It shows the dependence of the average grain
temperature on both the grain composition and 
size and on the radiation field
intensity.

Experimental studies show that molecular hydrogen formation on
amorphous 
silicate and carbon grains 
is efficient only in a narrow temperature window in the range
between 10-20 K
\cite{Pirronello1997,Katz1999,Perets2007}. 
This range coincides with the typical grain 
temperatures in diffuse clouds.
Therefore, such grains are efficient catalysts for H$_2$ formation 
in diffuse clouds, as long
as they are sufficiently large.
The efficiency of molecular hydrogen formation on small
grains is reduced under conditions in which the number
of adsorbed H atoms on the grain is of order 1 or lower
\citep{Tielens1982,Charnley1997,Caselli1998}.
Under these conditions, the rate equations fail
and stochastic methods based on the master equation
are needed in order to evaluate the reaction rates
\citep{Biham2001,Green2001,Charnley2001}.
Since the grain size distribution is dominated by small
grains, their effect on the rate of molecular hydrogen formation is
important.

The experimental results cannot 
explain the high abundance of H$_2$
in photon-dominated regions where the grain temperatures may reach
50 K
\cite{Habart2004}. 
This led to the possibility that temperature fluctuations of
small grains may enhance the formation rate of molecular hydrogen
and possibly resolve the discrepancy between the laboratory experiments
and astrophysical observations.

The effect of grain temperature fluctuations 
on the formation rate of
molecular hydrogen 
on small dust grains
in diffuse clouds
was studied recently
\citep{Cuppen2006}.
It was found that for radiation intensity characterized by
$\chi_{\rm MMP}=1$, 
molecular hydrogen formation is efficient
only on very rough grains which exhibit
higher energy barriers for H desorption
than those measured experimentally
\cite{Katz1999,Perets2007}. 
However, deeper in the cloud, where the radiation intensity 
is reduced by extinction,
molecular hydrogen formation is efficient 
also using the experimentally measured grain parameters.

To understand the effect of stochastic heating on hydrogen
recombination, consider a small grain exposed to
UV radiation.
During the time between the temperature spikes the grain
is colder than a large grain exposed to a similar radiation
field.
If the time between spikes is long enough for the
grain to adsorb several H atoms, these atoms
may quickly recombine during the next spike.
In this case, stochastic heating may enhance
H$_2$ formation on small grains.
However, 
even for $\chi_{\rm MMP}=1$
the rate of UV photon absorption on a grain is
much higher than the adsorption rate
of H atoms.
As a result, small grains rarely accumulate more than
one H atom between temperature spikes.
Thus, stochastic heating alone does not enhance the rate
of H$_2$ formation.

It was suggested that
chemisorption sites on grain surfaces may also enhance the
formation of H$_2$ in photon-dominated regions
\citep{Cazaux2002,Cazaux2004}.
However, experiments on H$_2$ 
formation on graphite indicate that
in order to enter the chemisorption 
sites the H atoms may need to
pass through an energy barrier
\citep{Zecho2002,Zecho2004}.
In addition, molecular hydrogen formation
from chemisorption sites is markedly different 
than the Langmuir-Hinshelwood mechanism observed
in physisorption sites.
In particular, H$_2$ molecules are formed only
upon desorption of the H atoms from the surface,
which occurs only at high surface temperatures
of $T>400$ K
\citep{Zecho2002,Zecho2004,Perets2006}.
These temperatures 
are much higher than the maximal grain temperatures
of stochastically-heated small grains 
in photon-dominated regions
(Fig. 6).
Therefore, it seems that even the combination of
stochastic heating and chemisorption does not account
for the formation of H$_2$ in photon-dominated regions.

\section{Summary}
\label{sec:dis_sum}

The temperatures of interstellar dust grains and 
their dependence on the grain size and composition 
and on the radiation field intensity have been 
studied using stochastic simulations.
It was found that grains of radii smaller than
0.02 $\mu$m exhibit temperature fluctuations which
become more intense as the grain size decreases.
The temperature distribution of small grains is
a skewed function, dominated by the low temperature 
side with a narrow tail in the high temperature side.
The fluctuations give rise to a reduction in
the average grain temperature. 
The average temperatures of grains of radii
larger than 0.035 $\mu$m are also slightly reduced
due to their more efficient emission of infrared
radiation.
The average grain temperatures vs. radiation intensity
were fitted according to
$\langle T \rangle = K {\chi_{\rm MMP}}^{\gamma}$,
where $K$ and $\gamma$
depend on the grain size and composition.
These results are more accurate and general than
the commonly used formulae for interstellar dust 
temperatures, which do not take into account
the detailed features of the absorption ane 
emission and the effects of stochastic heating.

\bibliographystyle{unsrt}

\end{document}